\documentclass[amsmath,notitlepage,aps,reprint,prl]{revtex4-1}

\usepackage{graphicx}
\usepackage[utf8]{inputenc}

\usepackage{array}
\usepackage{upgreek}
\usepackage{wasysym}
\newcommand{\dd}{\mathrm{d}}
\newcommand{\kB}{k_\mathrm{B}}
\newcommand{\ii}{\mathrm{i}}
\newcommand{\sub}[1]{_\mathrm{#1}}
\newcommand{\func}[2]{\mathrm{#1}\left(#2\right)}

\begin{document}

\title[]{Thermalization via Heat Radiation of an Individual Object Thinner\\ than the Thermal Wavelength}
\author{C. Wuttke}
\author{A. Rauschenbeutel}
\email{Arno.Rauschenbeutel@ati.ac.at.}
\affiliation{Vienna Center for Quantum Science and Technology, TU Wien – Atominstitut, Stadionallee 2, 1020 Wien, Austria.}
\date{\today}

\begin{abstract}
Modeling and investigating the thermalization of microscopic objects with arbitrary shape from first principles is of fundamental interest and may lead to technical applications. Here, we study, over a large temperature range, the thermalization dynamics due to far-field heat radiation of an individual, deterministically produced silica fiber with a predetermined shape and a diameter smaller than the thermal wavelength. The temperature change of the subwavelength-diameter fiber is determined through a measurement of its optical path length in conjunction with an ab initio thermodynamic model of the fiber structure. Our results show excellent agreement with a theoretical model that considers heat radiation as a volumetric effect and takes the emitter shape and size relative to the emission wavelength into account.
\end{abstract}

\maketitle
Thermalization via heat radiation is an omnipresent process which, e.g., determines the temperature of stars and planets or the functioning of incandescent lamps. For a perfectly black body, the spectral emissive power of far-field thermal radiation was first explained by Planck \cite{Planck1901} who used quantized energies for the radiation field, thereby breaking the grounds for quantum theory. 
The thermal radiation of a real object can then be related to that of a perfectly black body by introducing a correction factor, the so-called spectral emissivity which is treated as a surface property that depends on the specific material and the surface roughness \cite{Siegel1972}. However, as Planck already stated himself, his theory is designed to describe the far-field heat radiation of macroscopic bodies. Strictly speaking, it can therefore not be applied to objects that have a size or separation comparable to the thermal wavelength. In particular, as soon as the absorption length for the thermal spectrum gets comparable or larger than the size of the radiating body, thermal radiation becomes a volumetric effect and the spectral emissivity has to take the photonic properties of the object into account. These effects call for a more comprehensive theoretical description. Two established theoretical frameworks in this context are Mie scattering in combination with Kirchhoff's law and fluctuational electrodynamics (FED) \cite{Rytov1978,Bohren1998,Joulain2005,Carey2008}. Both approaches give accurate predictions for the far-field thermal radiation \cite{Carey2008} which, in case of the infinite cylinder, have been formally been shown to be identical \cite{Golyk2012}. In addition, the FED-framework has proven to be a versatile tool which also allows one to compute Casimir forces and the radiative heat transfer of arbitrarily shaped bodies in the far- and near-field \citep{Krueger2011}.

The radiative heat exchange of particles smaller than the thermal wavelength has been extensively studied in the past with ensembles of, e.g., soot particles or interstellar dust in the context of climate physics and astrophysics, respectively \cite{Bohren1998,Bond2006}. In general, these ensembles are not monodisperse, meaning that only statistical information on their size, shape, and material properties is available. More recently, the progress in nanofabrication led to a new approach using deterministically produced samples of well-defined shape and material. This way, the effect of size and geometrical structure of an object on the spectrum, coherence, and angular distribution of its far-field thermal radiation has been investigated \citep{Greffet2002,Fan2009,Schuller2009,Shen2009,Liu2011b}. Moreover, it has been shown that the radiative heat transfer rate can be strongly enhanced by near-field effects \cite{Rousseau2009}. 

\begin{figure}
	\includegraphics[width=8.5cm]{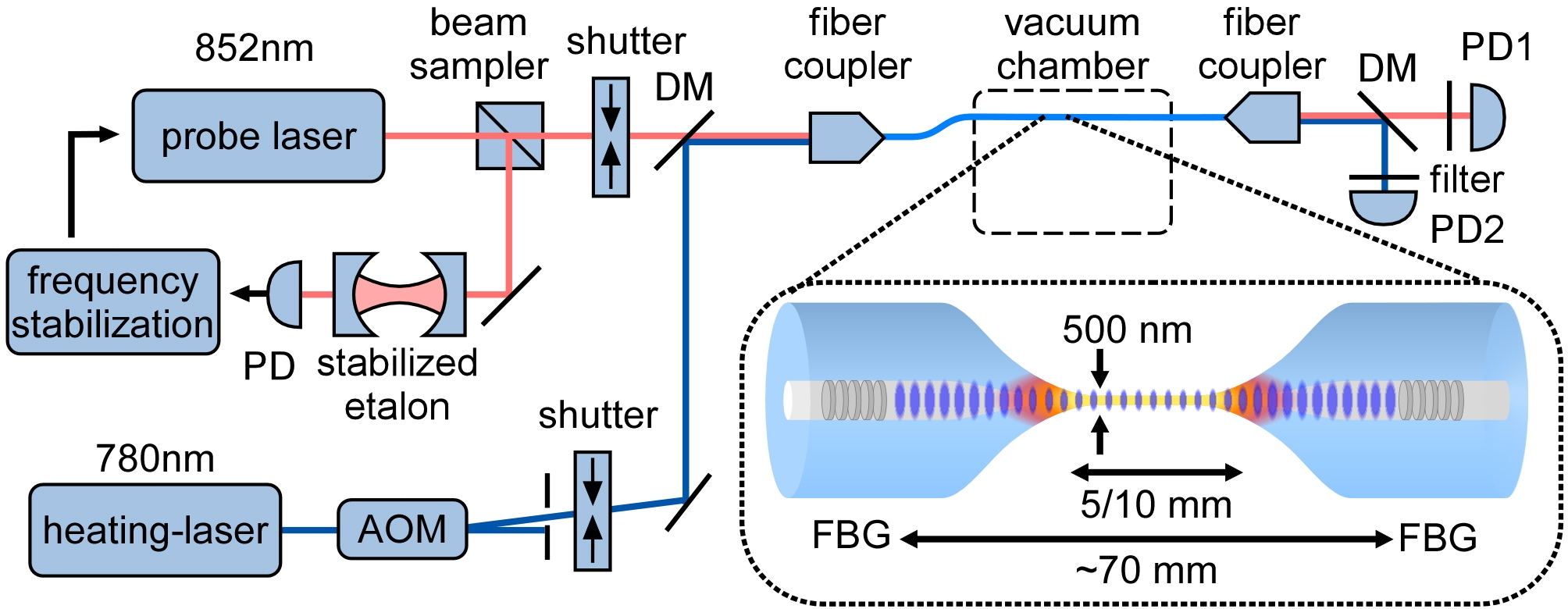}
	\caption{Schematics of the experimental setup for measuring the thermally induced optical path length change of the optical nanofiber. The inset shows the TOF-resonator with the nanofiber waist.\label{fig1}}
\end{figure}

Here, we measure the thermalization dynamics of an individual object of predetermined size and shape which is thinner than the thermal wavelength and all absorption lengths of the relevant part of the thermal radiation spectrum. For this purpose, we employ a silica nanofiber with a diameter of 500~nm and monitor its temperature-dependent optical path length interferometrically during heating and cooling. The fiber resides in an ultra high vacuum chamber and heat transport is dominated by far-field heat radiation. Our data show that the total far-field radiated power of this sub-wavelength structure agrees quantitatively with ab-initio predictions given by FED \cite{Golyk2012} for temperatures ranging from room temperature up to and beyond the glass transition temperature of fused silica. 

\begin{figure*}[htb]
	\includegraphics[width=16.2cm]{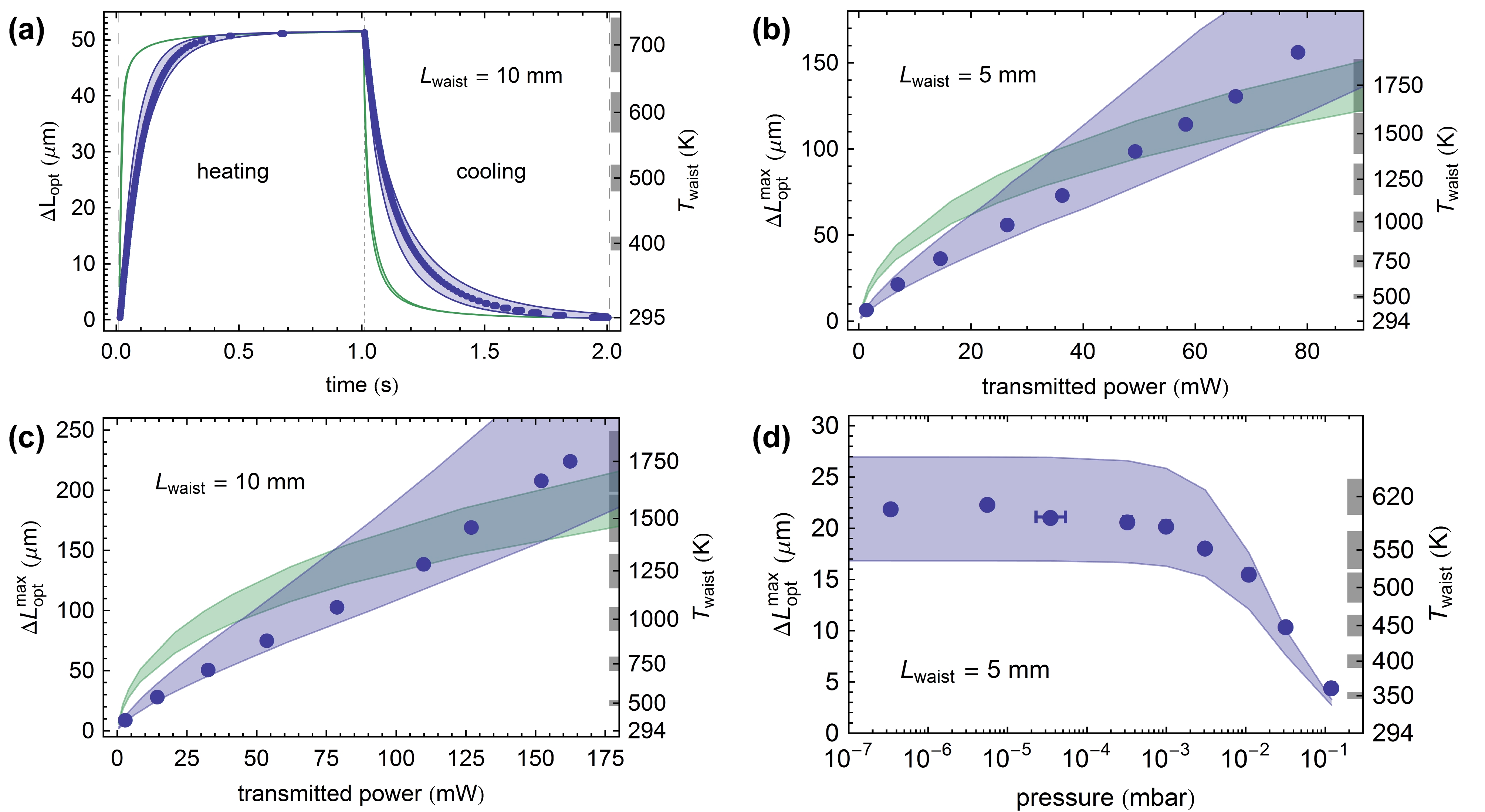}
	\caption{Thermally induced optical path length change of the silica nanofiber. Circles: Experimental data; blue bands: FED predictions; green bands: predictions using Planck’s law. (a) Example of thermalization dynamics. (b-d) Maximum optical path length change as a function of the heating laser power transmitted through the TOF and background gas pressure, respectively. \label{fig2}}
\end{figure*}
The silica nanofibers used for our experiment are realized as the waist of a tapered optical fiber (TOF) which is produced from a standard optical fiber in a heat-and-pull process \cite{Birks1992, Warken2008}. A temperature change of the nanofiber waist induces a change of its optical path length which is primarily due to the thermo-optic effect of silica. In order to measure this optical path length change, the TOF is enclosed in a fiber-based Fabry-P\'erot-type optical resonator which uses two fiber Bragg-gratings (FBGs) as cavity mirrors \cite{Wuttke2012}, see inset of Fig.~1. 

In this configuration, the optical path length change is translated into a frequency shift of the Fabry-P\'erot resonances which we read out via the transmission of a probe laser field, see Fig.~1. The nanofiber waist is heated by sending a second laser field through the TOF. Its wavelength lies in the transmission band of the FBGs and its power is adjustable by means of an acousto-optic modulator (AOM). A small fraction of this optical power is absorbed along the nanofiber, thereby heating the latter. After exiting the TOF, the transmitted probe and heating laser fields are separated by a dichroic mirror (DM) and spectral filters such that their respective powers can be measured independently using photodiodes (PD1 \& PD2). By abruptly switching the heating laser beam with a mechanical shutter, the heat source can be turned on and off, allowing us to measure the heating and cooling dynamics of the nanofiber. 

For this purpose, we record a time trace of the probe transmission which consists of a sequence of Fabry-P\'erot transmission peaks. From one peak to the next, the single-pass intra-cavity optical path length has changed by half the vacuum wavelength of the probe laser, $\Delta L_\mathrm{opt}=\lambda_0/2=426$~nm. A typical time trace of $\Delta L_\mathrm{opt}$ is shown in Fig.~2a for a TOF-resonator with a waist diameter of (500$\pm$50)~nm, a nominal homogeneous waist length of $L_\mathrm{waist}=10$~mm, and a finesse of $85\pm1$ (TOF-resonator \#1; see supplementary material (SM) for a detailed radius profile). 
For this measurement, the heating laser is switched on between $t=0...1$~s, the transmitted power is $P_\mathrm{heat}=32.7(1)$~mW, and the background gas pressure is $p=10^{-6}$~mbar. For both the heating and cooling process, $\Delta L_\mathrm{opt}$ varies rapidly during the first 500~ms. Following this initial fast dynamics, the variation of $\Delta L_\mathrm{opt}$ takes place on a much longer timescale on the order of several seconds. Because of the discrete nature of the measurement method which relies on counting transmission peaks, one additional data point was added at the end of the heating (cooling) period. Its value is chosen $\lambda_0/4$ higher (lower) than the last measured data point and its error bar is given by $\pm\lambda_0/4$, thereby accounting for the fact that $\Delta L_\mathrm{opt}$ changed by less than $\lambda_0/2$. 

In order to gain quantitative understanding of the observed dynamics of $\Delta L_\mathrm{opt}$, we develop an ab initio thermodynamical model of the TOF including heat transport via thermal radiation and heat conduction along the TOF. We experimentally checked that heat diffusion through the surrounding background gas is negligible for pressures below $10^{-4}$~mbar (see below). Moreover, we assume that the fiber is thermalized across its cross-section. The system can then be described by the following differential equation in units of power \cite{Demtroeder1}:
\begin{multline}
		c\sub{p}\,\rho\,\partial_t T\,\dd V = - \dd H\sub{rad}(T) + \dd H\sub{rad}(T_0)\\
		 + \nabla(\lambda \nabla T)\,\dd V+ \dd P\sub{heating} ~, \label{eq:heat:ode:3d} 
\end{multline}
where $c\sub{p}=(700-1500)$~J/(kg$\cdot$K) and $\rho=2200$~kg/m$^3$ are the specific heat and density of silica \cite{Horbach1999,Freeman1986}, $\dd P\sub{heating}$ is a heat source, $\lambda = (1.3-2.1)$~W/(K$\cdot$m) is the heat conductivity of silica \cite{Freeman1986}, $T_0=(294\pm0.5)$~K is the room temperature. The dependencies on space and time are omitted for clarity. For comparison, the radiated heat $\dd H_\mathrm{rad}$ is calculated using both, a na\"{i}ve approach given by Planck’s law in combination with the spectral emissivity of a silica$-$vacuum interface (see SM) and the predictions from FED calculations as computed in \cite{Golyk2012} (details in the SM).

It has been shown that the transmission loss of silica nanofibers is primarily governed by surface pollution \cite{Fujiwara2011b}. In our model, we therefore assume that the heating of the TOF is caused by surface absorption. We note, however, that our analysis only weakly depends on this assumption and that the other extreme case of heating by pure volume absorption yields practically the same results. Despite the fact that the absorption of the heating laser is dominated by pollutants, the overall emittance of the fiber is well-described by assuming a pure silica fiber (see SM): In essence, the absorption of the pollutant in the near infrared (NIR: $\sim(0.7$--$2)~\upmu$m) spectral range only dominates because of the extreme transparency of the silica. However, it is still extremely small and, in particular, negligible compared to that of silica in the wavelength range where the latter is opaque. As a result, the emittance of the total structure (silica nanofiber including the pollutants) is dominated by the properties of silica in the relevant spectral regions, i.e., where the vast majority of the heat radiation is emitted.

The largest uncertainties that enter into the model are the radius profile, which is known with a relative error of $\pm 10~\%$ \cite{Wiedemann2010,Stiebeiner2010}, and the wavelength-dependent complex refractive index of fused silica, $\hat{n}$. The latter is the sole material property that is used in the FED calculation of the radiated power as well as in the calculation of the spectral emissivity. It was extracted from various literature sources for wavelengths ranging from $\approx 30$~nm to $\approx 2$~mm (see SM).

We numerically solve the two resulting differential equations (Planck and FED) for the same heating and cooling cycle as in the experiment and determine the time-dependent temperature profile along the TOF as a function of the total absorbed optical power, $P_\mathrm{abs}$. Four different parameter sets are used for these calculations, combining the minimum and maximum TOF radius profile with the values of $\hat{n}$ that yield the minimum or maximum radiated power per unit length, respectively. From these temperature profiles, we then compute $\Delta L_\mathrm{opt}$ via the temperature-dependent effective refractive index which is integrated along the TOF-resonator (see SM). The theory bands in Fig.~2a are delimited by the two extremal time traces of $\Delta L_\mathrm{opt}$ that reproduce the maximum observed optical path length change $\Delta L_\mathrm{opt}^\mathrm{max}$ (solid lines). We find excellent agreement between the time dependence of $\Delta L_\mathrm{opt}$ predicted by FED and the experimental data while the prediction for Planck’s law predicts a considerably faster thermalization.

Figure 2b and c show $\Delta L_\mathrm{opt}^\mathrm{max}$ as a function of the heating laser power transmitted through the TOF, $P_\mathrm{heat}$, for TOF-resonators \#1 and \#2. The latter corresponds to a TOF with a waist diameter of $(500\pm50)$~nm, $L_\mathrm{waist}=5$~mm, and a finesse of $18\pm 1$. In both cases, $\Delta L_\mathrm{opt}^\mathrm{max}$ increases roughly linearly with $P_\mathrm{heat}$ within the observed power range. In order to analyze the measured data, the ratio $\eta$ between $P_\mathrm{heat}$ and $P_\mathrm{abs}$ has to be determined. This is done by inserting $P_\mathrm{abs}=\eta\,P_\mathrm{heat}$ into both theories and fitting their predictions independently to the data using the four parameter sets concerning the radius and the complex refractive index. We compute the mean of the minimum and maximum resulting $\eta$-value, $\bar{\eta}=(\eta_\mathrm{max}+\eta_\mathrm{min})/2\approx 2$~\permil~for FED and $\bar{\eta}\approx 3$~\% for Planck's law and plot the extremal theory prediction (solid lines bounding the colored bands) using the corresponding two parameter sets and $P_\mathrm{abs}=\bar{\eta}\, P_\mathrm{heat}$. For both resonators, the FED prediction is in excellent agreement with the experimental data while Planck’s law predicts a sub-linear increase of the equilibrium temperature, and therefore of $\Delta L_\mathrm{opt}^\mathrm{max}$. This qualitative deviation stems from the different scaling of the radiated power with temperature for the two models: $T^4$ assuming Stefan-Boltzmann law and $< T^3$ in the case of the FED prediction.

From the temperature profiles along the TOF which lead to the two extremal predictions of $\Delta L_\mathrm{opt}^\mathrm{max}$, we determine two corresponding extremal predictions for the temperature at the center of the waist, $T_\mathrm{waist}^\mathrm{max}$ and $T_\mathrm{waist}^\mathrm{min}$, respectively. The resulting non-linear temperature scale, $\bar{T}_\mathrm{waist}=(T_\mathrm{waist}^\mathrm{max}+T_\mathrm{waist}^\mathrm{min})/2$, where $\Delta\bar{T}_\mathrm{waist}=(T_\mathrm{waist}^\mathrm{max}-T_\mathrm{waist}^\mathrm{min})/2$, is shown on the right axes of all panels of Fig.~2, with the ticks and the gray bars indicating $\bar{T}_\mathrm{waist}$ and $\Delta\bar{T}_\mathrm{waist}$, respectively. It covers a temperature range from room temperature to almost 2000~K. We note that the temperature scale in Fig.~2a refers to the cooling process. The corresponding scale for the heating process (not shown for clarity) differs from the latter because two distinct spatial temperature profiles arise during heating and cooling. As a result, the same optical path length change corresponds to two different waist temperatures.

\begin{figure*}
	\includegraphics[width=15.8cm]{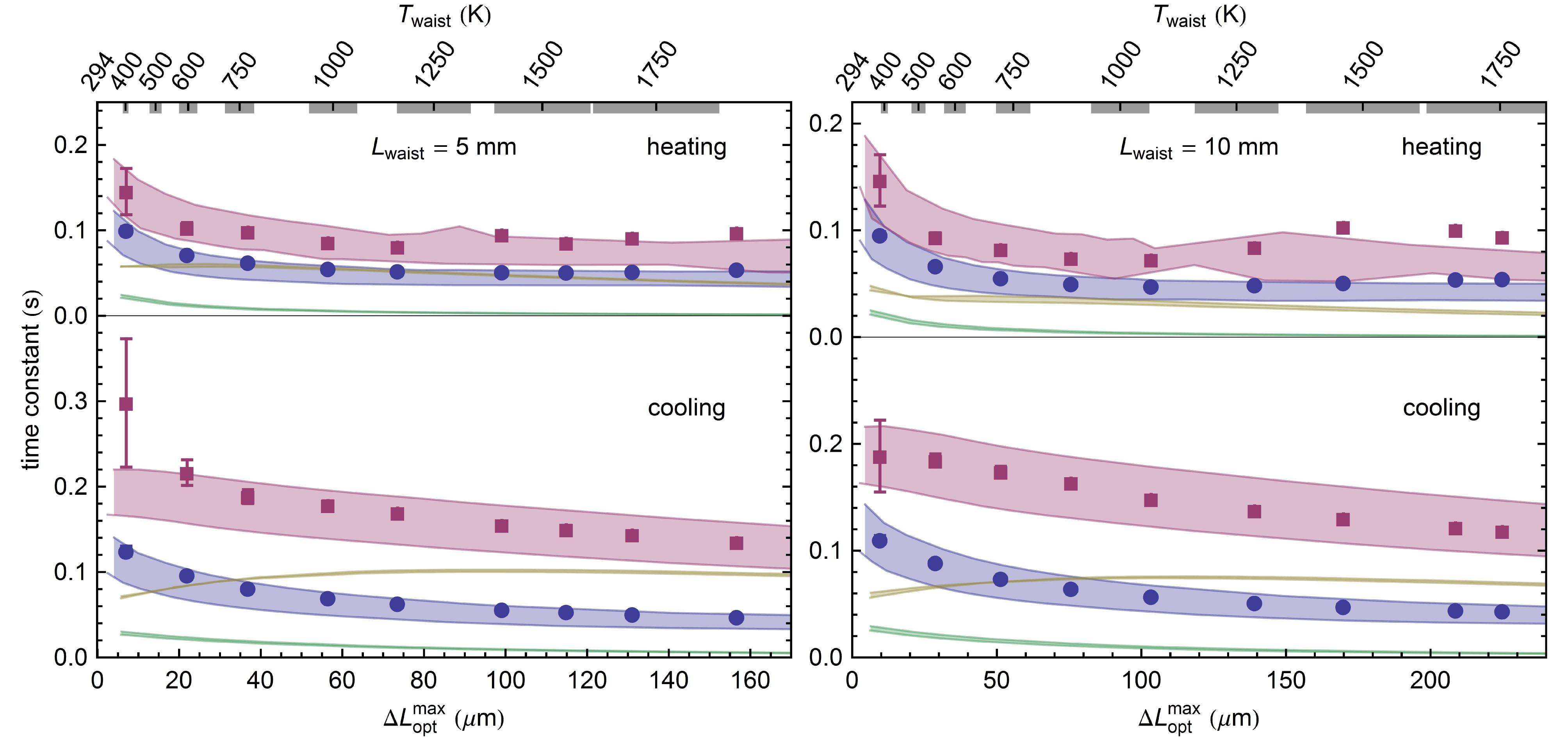}
	\caption{Thermalization time constants of the silica nanofiber as a function of the maximum optical path length change for the two TOF resonators. Blue circles: measured initial time constants; red squares: final time constants; blue (red) bands: FED predictions for initial (final) time constants; green (yellow) bands: predictions for initial (final) time constants using Planck’s law. \label{fig3}}
\end{figure*}
While the upper temperature limit exceeds the glass transition temperature of fused silica of about 1450~K \citep{Doremus2002}, the nanofiber did not fuse. In an additional measurement (see SM) the fiber survived temperatures of up to $(2515\pm 255)$~K. A theoretical value for this temperature can be obtained by considering the fiber as a filament with a temperature-dependent viscosity \cite{Koulakis2008,Doremus2002}. We find the melting dynamics to take place at a comparable time scale as the experiment, meaning that the fiber would actually break during the experimental cycle, at a temperature of $(2710\pm140)$~K which is in good agreement with the experimental result. Based on similar considerations, we also confirmed that the fiber will maintain a minimum stress of $\sim 100$~kPa during the experiments if the maximum temperature is $\sim 1800$~K (see SM).

Figure 2d shows $\Delta L_\mathrm{opt}^\mathrm{max}$ measured as a function of the background gas pressure for TOF-resonator \#2 and $P_\mathrm{heat}=13.2(4)$~mW. The theory prediction takes heat diffusion through the background gas into account while neglecting temperature gradients. In our experiment, this treatment is valid for pressures smaller than $10^{-3}$~mbar and overestimates the gas-based heat transport for higher pressures (see SM). We find that $\Delta L_\mathrm{opt}^\mathrm{max}$ is constant for pressures lower than $10^{-4}$~mbar, confirming that heat diffusion through the background gas is negligible for all other measurements presented here, which were performed at pressures below $10^{-6}$~mbar.

We now analyze the thermalization dynamics for varying heating powers and thus for varying $\Delta L_\mathrm{opt}^\mathrm{max}$. For this purpose, we determine the $10~\%-50~\%$ and $75~\%-90~\%$ rise times (heating) and the $90~\%-50~\%$ and $25~\%-10~\%$ fall times (cooling) of the time traces of $\Delta L_\mathrm{opt}$ in order to quantify the initial and the final dynamical behavior. We find time constants on the order of 100~ms which are plotted in Fig.~3 as a function of $\Delta L_\mathrm{opt}^\mathrm{max}$.
The same quantities are derived from the theory predictions for Planck’s law and FED, respectively, and plotted as bands in the same graphs. The agreement between the experimental data and the FED predictions is excellent, whereas Planck’s law predicts up to one order of magnitude shorter time constants and, therefore, highly overestimates the radiated power from the nanofiber. We note, that all theory predictions in this figure are ab initio results without any adjustable parameters.

Summarizing, using a silica nanofiber of predetermined shape with a diameter smaller than the thermal wavelength, we measured the thermalization dynamics of an individual nanoscopic object over a large temperature range. Our analysis confirms that the total thermally radiated power in the far-field is accurately predicted over a large temperature range using an ab initio thermodynamical model based on fluctuational electrodynamics, material properties, and the size and shape of the emitter. Modeling the thermalization of microscopic particles with arbitrary shape from first principles has important applications in the framework of, e.g., heat management in nano-devices, radiative forcing of aerosoles in the earth's atmosphere \cite{Bond2013}, or cavity opto-mechanics experiments \cite{Chang2010}.

We thank M. Rothhardt and co-workers for the FBGs, M. Ponweiser for devising the thermal initialization method, and V. Golyk and M. Kr\"uger for the FED implementation. We acknowledge financial support by the Volkswagen Foundation (Lichtenberg Professorship), the ESF (EURYI Award), NanoSci-E+ (NOIs project),  and the Austrian Science Fund (FWF; SFB FoQuS Project No. F 4017).

\clearpage
\section*{Supplementary material}
\subsection*{Details for the thermodynamical model}
We model the system including heat transport via radiation, heat conduction in silica, and heat diffusion through the surrounding gas. The assumption that no temperature gradients appear in the surrounding background gas is valid for mean free path lengths that are much larger than the distance from the fiber to the heat sink, such as, e.g., the fiber mount or the vacuum chamber wall. We estimate the pressure up to which this condition is fulfilled by assuming a typical distance to the heat sink in our system of 10~cm and thereby obtain $p<10^{-3}$~mbar. We note, however, that the transition to the regime in which the background gas exhibits a temperature gradient can take place in a pressure range that covers several orders of magnitude. For this reason, we also performed an experimental check of the validity of our assumption, see Fig.~2d in the article. The system can then be described by the following differential equation in units of power \cite{Demtroeder1}:
\begin{multline}
		c\sub{p}\,\rho\,\partial_t T\,\dd V = - \dd H\sub{rad}(T) +\dd H\sub{rad}(T_0) + \dd P\sub{heating}\\
		+\nabla(\lambda \nabla T)\,\dd V -  
		\sqrt{\frac{k_\mathrm{B}\,f^2}{8\pi\,M\,T_0 }} p\, (T-T_{0})\,\dd S  \label{eq:heat:ode:3d} 
\end{multline}
where $\dd P\sub{heating}$ is a heat source, the remaining quantities are given in Tab.~\ref{tab:heat:ode:parameters}, and the dependencies on space and time are omitted for clarity.
\begin{table*}[htb]
	\centering
	\begin{tabular}{|c|c|>{\centering}m{4cm}|>{\centering}m{3cm}|c|}
		\hline
		symbol & description & value from $T_0$ to the highest temperature covered in Ref.& temperature range convered in Ref.& Ref. \\\hline
		$T_0$ & room temperature & $(294\pm0.5)$~K & - & - \\\hline
		$\lambda$ & heat conductivity & $(1.3-2.1)$~W/(K$\cdot$m)&$(0.1-1650)$~K& \cite{Freeman1986}\\\hline
		$c\sub{p}$ & specific heat (silica)& $(700-1500)$~J/(kg$\cdot$K)&  $(0-2000)$~K&\cite{Horbach1999}\\\hline
		$\rho$ & density (silica) & $2200$~kg/m$^3$ & - & \cite{Freeman1986} \\\hline
		$f$ & degrees of freedom of N$_2$& $6$ & - & - \\\hline
		$M$ & molecular mass of N$_2$ & $4.653\cdot 10^{-26}$~kg & - & -\\\hline
		$k_\mathrm{B}$ & Boltzmann's constant & $1.381~\cdot 10^-{-23}$~J/K & - &-\\\hline
	\end{tabular} 
	\caption{Model parameters and their values with references.}\label{tab:heat:ode:parameters}
\end{table*}
The radiated heat $\dd H_\mathrm{rad}$ can be either calculated using Planck’s law in combination with the spectral emissivity of a silica$-$vacuum interface (see below) or using the predictions from FED calculations \cite{Krueger2011,Golyk2012}.

We assume the fiber is heated via surface absorption so that the local heating power per unit length is given by the local surface intensity $I(a(z))$ of the fiber mode multiplied by the fiber circumference $\partial_z P_\mathrm{heating}(z)=2\pi\,a(z)\,I(a(z))$. Due to the small diameter of the nanofiber, we assume that it exhibits a constant temperature across its diameter. Moreover, the fiber exchanges an extremely small amount of heat with the vacuum chamber which, therefore, stays at room temperature during the experiment. This allows us to reduce the equation to one dimension in space along the fiber axis ($z$-axis). After differentiating the resulting equation with respect to $z$, we then find the following differential equation in units of power per length : 
\begin{multline}
\pi a^2\,c\sub{p}\,\rho\,\partial_t T = - \partial_z H\sub{rad}(T)+\partial_z H\sub{rad}(T_0)  + \partial_z P\sub{heating}\\
+\pi a^2\partial_z\left(\lambda\,\partial_z T \right)- \sqrt{\frac{k_\mathrm{B}\,f^2}{8\pi\,M\,T_0}} \,p\,(T-T_{0})\,2\pi a \label{eq:heat:ode:1d}
\end{multline}  
where $\partial H_\mathrm{rad}$ describes the radiated power per unit length. This equation is solved numerically for a given radius profile using the method of lines implemented in Mathematica with a spatial discretization of 0.1~mm along the fiber axis for a TOF section that extends at least 1~cm beyond either side of the waist: $L_\mathrm{sim}=L_\mathrm{waist}+2$~cm. At these points, the fiber diameter has increased to $\sim$10~$\mu$m and is thus a factor of 20 larger than the waist diameter. As a consequence, the evanescent field and hence the heating power vanishes and the heat capacity per unit length increases by a factor of 400. The boundary conditions are chosen such that the fiber is thermalized to ambient temperature at the beginning of the simulation $T(t=0,z)=T_0$ and no heat flow occurs at the boundary of the simulation cell: $\partial_z T(t,\pm L_\mathrm{sim}/2)=0$. By solving equation~\ref{eq:heat:ode:1d}, we obtain a temperature profile as a function of time and position along the fiber, $T(t,z)$. We then use the latter to compute the optical path-length for discrete time-steps via the thermally induced effective refractive index change (see below) and compare the results from this simulation to the data. 

\subsection*{Complex refractive index of fused silica}
The complex refractive index, $\hat{n}=n+ik$, is the sole material property that enters the calculation of the spectral emissive power of thermal radiation. We take the uncertainty of the literature values into account by using all four combinations of the maximum and minimum of each $n$ (refractive index) and $k$ (extinction coefficient), respectively, in order to compute the total radiated power. We then use the maximum and minimum value of the latter for our further calculations.

Optical fibers are made of high-quality synthetic fused silica, which is an amorphous glass, with a core with typically a few weight percent of dopants such as, e.g., germanium. Four types of fused silica are commercially available \citep{Kitamura2007}: Type I and Type II are produced from crystal quartz by electric melting and from quartz powder by flame fusion, respectively, whereas Type III and Type IV glasses are synthetic glasses produced from SiCl$_4$ in a H$_2$-O$_2$ flame and in a water-vapor free plasma flame respectively. The synthetically produced glasses feature the lowest impurity content and, thereby, exhibit a higher transmission than the Type I and Type II glasses, in particular in the near-infrared and in the ultraviolet spectral ranges. 
	
\begin{figure}[h]
	\centering
	\includegraphics[width=8.4cm]{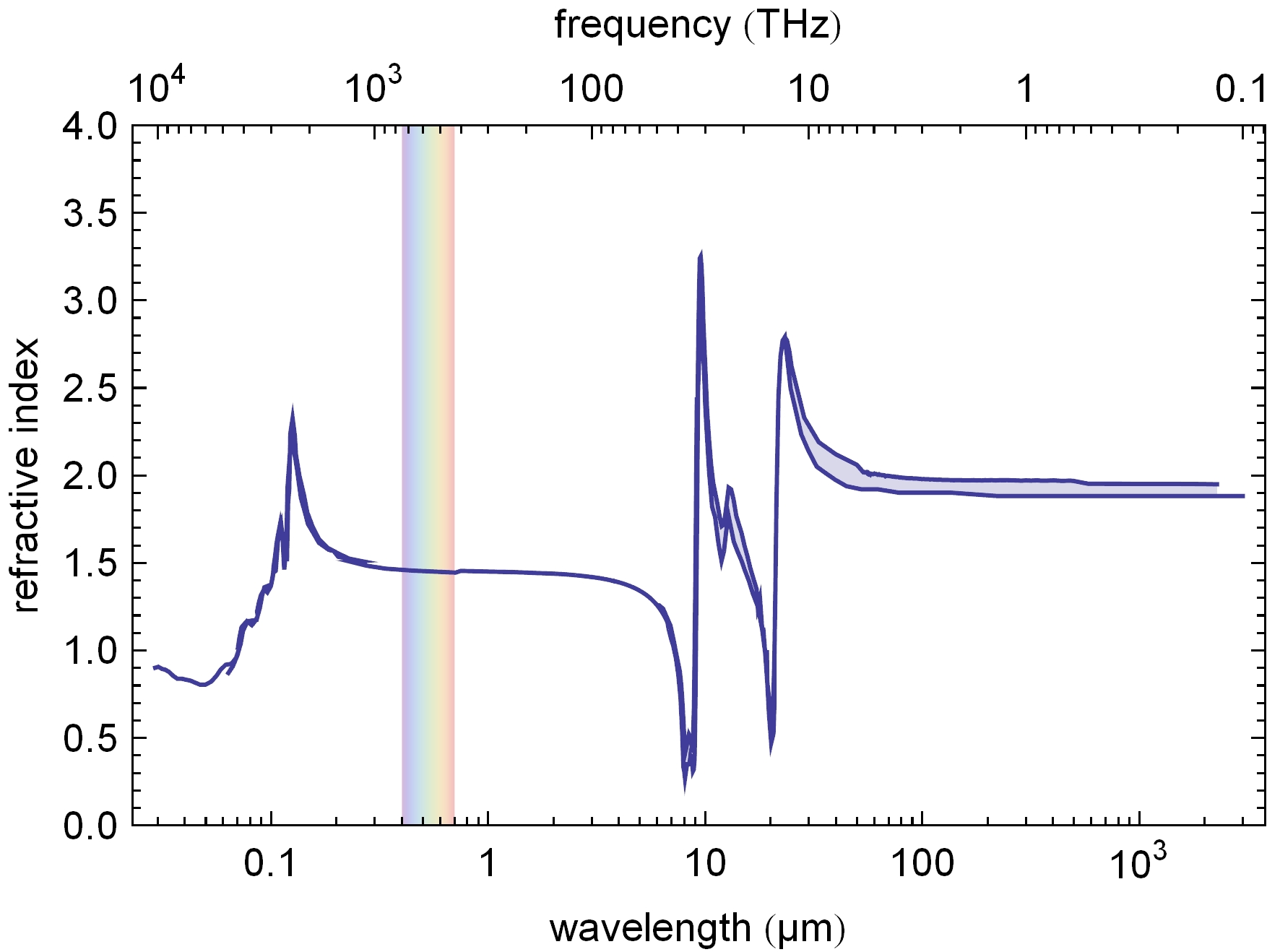}\vspace{2mm}
	\includegraphics[width=8.4cm]{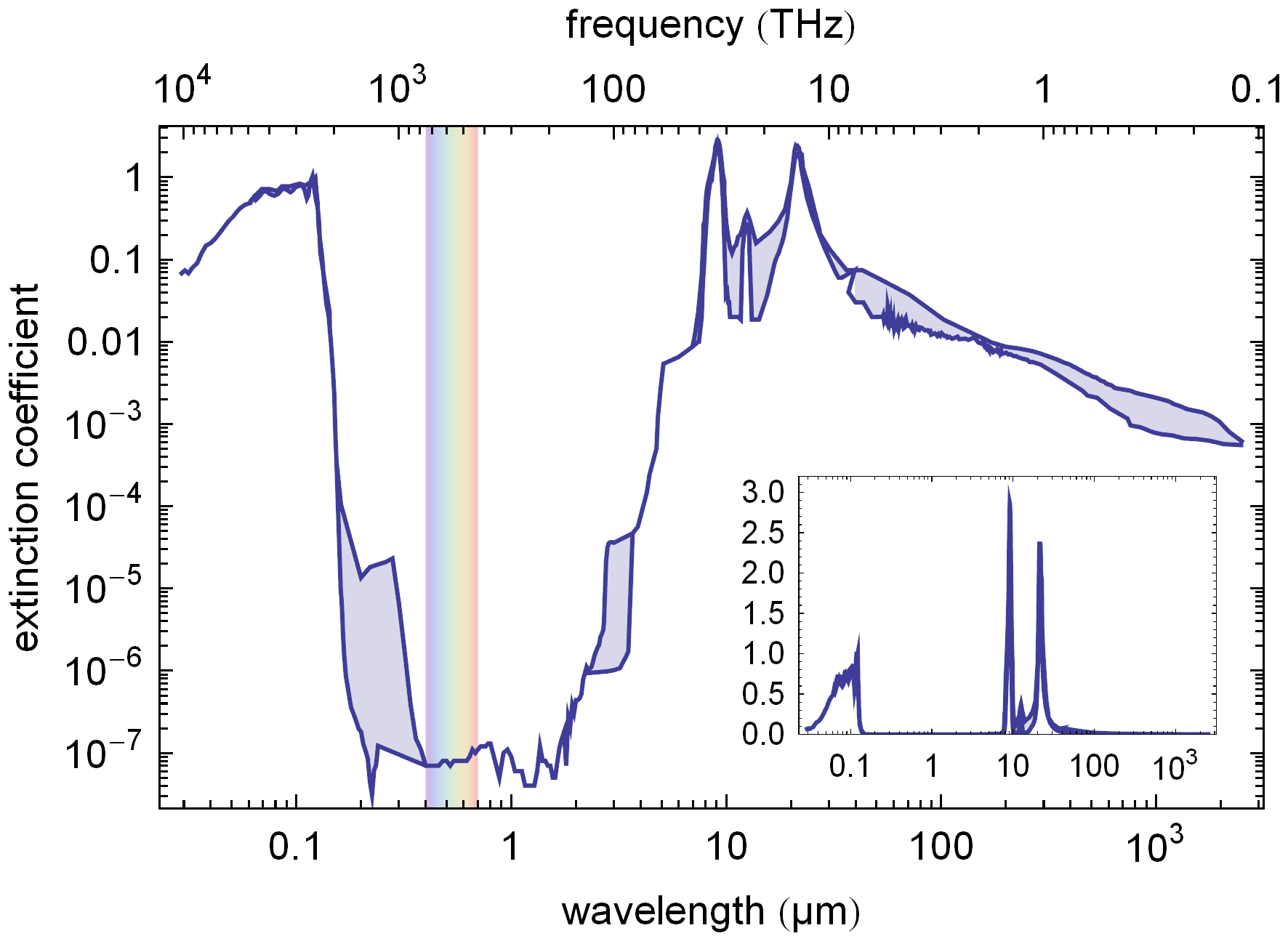}
	\caption{Refractive index and extinction coefficient of fused silica as function of the wavelength. Red curve: maximum value; blue curve: minimum value.}\label{fig:nhat:silica}
\end{figure}

Since our TOF fabrication employs a H$_2$-O$_2$ flame for the tapering process, we consider the glass in the TOF to be a mixture of Type III and Type IV silica. We therefore extract the minimum and maximum values of both $n$ and $k$, respectively, that are available for these two types of glass for our calculations. For wavelength regions in which we did not encounter literature values for these particular types of glass, we also use data for other types of fused silica. The resulting minimum and maximum contributions of the complex refractive index that we employ for our calculations were derived from \citep{Afsar1984,Grischkowsky1990,Naftaly2007,Cunningham2010,Kitamura2007} and are shown in Fig.~\ref{fig:nhat:silica} as function of the optical wavelength.

We note that the values given for the extinction coefficient in the visible spectrum are at the lower limit of the parameter range that is experimentally accessible and therefore can be considered as maximum values. Since no further data is available and heat radiation in the visible spectrum contributes only marginally to the total heat transfer rate, we do not consider smaller values.

\subsection*{Far-field heat radiation of a cylinder}
In this section we resume the equations used to compute the far-field heat radiation of a cylinder as derived in \citep{Golyk2012}. The radiated power per unit length, $\partial_z H\sub{rad}$, of an infinite cylinder of radius $a$ at temperature $T\sub{c}$ that is composed of an isotropic material with the spectral dielectric function $\epsilon$ and spectral magnetic permeability $\mu$ can be expressed as 
\begin{multline}
	\partial_z H\sub{rad} =\frac{4}{c_0} \int_0^\infty
 \dd\nu\; \frac{h\,\nu^2}{\func{exp}{h\nu/\kB T\sub{c}}-1}
 \sum_{P=\parallel,\perp}\;\sum_{l=-\infty}^{\infty}\\
 \times \int_{-1}^{1}\dd \xi\;\left(\func{Re}{T_{l,\xi}^{PP}}+|T_{l,\xi}^{PP}|^2+|T_{l,\xi}^{P\bar{P}}|^2\right) ~,\label{eq:therm:rad:cylinder}
\end{multline}
where $\nu$ denotes the optical frequency, $c_0$ is the speed of light in vacuum, $P=\{\perp,\parallel\}$ denotes the polarization of the radiated light with respect to the plane given by the propagation direction of the emitted radiation and the cylinder axis (z-axis), $\bar{P}$ is the polarization perpendicular to $P$, $l$ is the mode number, $k_0=2\pi\,\nu/c_0$ is the vacuum propagation constant, and $\xi = k_z/k_0$ ratio of the axial propagation constant ($k_z$) and the vacuum propagation constant, and therefore the cosine of the emission angle with respect to the fiber axis $\xi = \cos(\theta)$ with $\theta \in [0,\pi]$. 

The T-matrix elements are given by
\begin{gather*}
	T_{l,\xi}^{\perp\,\perp} = -\,\frac{J_l(q\,a)}{H^{(1)}_l (q\,a)}\,\frac{\Delta_1\Delta_4-K^2}{\Delta_1\Delta_2-K^2}~,\\
	T_{l,\xi}^{\parallel\,\parallel} = -\,\frac{J_l(q\,a)}{H^{(1)}_l (q\,a)}\,\frac{\Delta_2\Delta_3-K^2}{\Delta_1\Delta_2-K^2}~,\\
	T_{l,\xi}^{\perp\,\parallel}=T_{l,\xi}^{\parallel\,\perp} = \frac{2\ii\,K}{\pi\,\sqrt{\epsilon\mu}\,(q\,a\,H_l^{(1)}(q\,a))^2}\,\frac{1}{\Delta_1\Delta_2-K^2}~,
\end{gather*}
where $J_l(x)$ is the Bessel function, $H_l^{(1)}(x)$ is the Hankel function of the first kind and the prime denotes the derivative with respect to the argument: $f'(x)=\partial_x f(x)$. The variable $q=k_0 \sqrt{1-\xi^2}$ denotes the component of the propagation constant that is perpendicular to the cylinder axis. Furthermore the following definitions have been used:
\begin{align*}
	\Delta_1 &= \frac{J_l'(q_1\,a)}{q_1\,a\,J_l(q_1\,a)}-\frac{1}{\epsilon}\frac{H_l^{(1)\prime}(q\,a)}{q\,a\,H_l^{(1)}(q\,a)}~,\\
	\Delta_2 &= \frac{J_l'(q_1\,a)}{q_1\,a\,J_l(q_1\,a)}-\frac{1}{\mu}\frac{H_l^{(1)\prime}(q\,a)}{q\,a\,H_l^{(1)}(q\,a)}~,\\
	\Delta_3 &= \frac{J_l'(q_1\,a)}{q_1\,a\,J_l(q_1\,a)}-\frac{1}{\epsilon}\frac{J_l'(q\,a)}{q\,a\,J_l(q\,a)}~,\\
	\Delta_4 &= \frac{J_l'(q_1\,a)}{q_1\,a\,J_l(q_1\,a)}-\frac{1}{\mu}\frac{J_l'(q\,a)}{q\,a\,J_l(q\,a)}~,\\
	K&=\frac{l\,\xi\,k_0\,c_0}{\sqrt{\epsilon\mu}\,R^2\,\omega}\left(\frac{1}{q_1^2}-\frac{1}{q^2}\right)~\mathrm{,~and }\\
	q_1&=k_0\sqrt{\epsilon\mu\,-\xi^2}~.
\end{align*}
The material properties enter solely via the dielectric function and permeability which is given above. We note that these results are the same as the ones obtained from Mie theory.

From Eq.~\ref{eq:therm:rad:cylinder} we now compute the spectral emissivity $\epsilon(\nu)$ of a silica cylinder by normalizing the total radiated power per surface area of the fiber $\partial_x\partial_\omega H\sub{rad}/(2\pi a)$ to the spectral emissive power of a black body given by Planck's formula \citep{Lienhard2011}:
\begin{align}
	P_\nu(\nu,T)\;\dd\nu&= \int \dd \Omega\;\cos(\theta)\, W(\nu,T)  \notag\\
	&=\frac{2\pi\nu^2}{c_0^2}\,\frac{h\nu}{\exp{(h\nu/k_\mathrm{b}T)}-1}\;\dd\nu~,\label{eq:Plancks:law}
\end{align}
where we have already integrated the spectral radiance, $W(\nu,T)$, over all emission angles in the hemisphere, $\dd \Omega =\smallint_0^{2\pi}\dd\phi \smallint_0^{\pi/2}\dd\theta\,\sin(\theta)$, while simultaneously considering the geometry factor: $\smallint\dd\Omega\;\cos(\theta)=\pi$. Then, the total radiated power of the fiber can be computed using:
\begin{equation}
H\sub{rad}(T) =S\sub{cyl}\,\int_0^\infty \dd\nu\,P(\nu,T)\,\epsilon(\nu)~,
\end{equation}
where $S\sub{cyl}=2\pi\,a\,\Delta z$ is the surface area of the cylinder. From Eq.~\ref{eq:therm:rad:cylinder} and Eq.~\ref{eq:Plancks:law} we obtain the following expression for the spectral hemispherical emissivity:
\begin{multline}
	\epsilon(\nu) = \frac{c_0}{\nu\,\pi^2\,a}\;
	\sum_{P=\parallel,\perp}\;\sum_{n=-\infty}^{\infty}\int_{-1}^{1}\dd \xi\;\\
	\times\left(\func{Re}{T_{l,\xi}^{PP}}+|T_{l,\xi}^{PP}|^2+|T_{l,\xi}^{P\bar{P}}|^2\right) ~.
	\label{eq:emissivity:cylinder}
\end{multline}
We would like to stress that this spectral emissivity is only valid for the particular fiber diameter it has been computed for and yields incorrect results when used for different diameters by scaling the surface area, as in standard heat radiation theory. In particular, the heat radiation of very thin cylinders scales with the volume. This illustrates that the concept of emissivity as material and surface property cannot be applied to subwavelength objects. However, a structure-dependent emissivity is a useful tool for the numerical computation of the radiated power because the costly numerically integration of the T-matrix only needs to be evaluated once for each diameter and then may be used to compute $H\sub{rad}$ for several temperatures. 

\begin{figure}
	\centering
	\includegraphics[width=8.4cm]{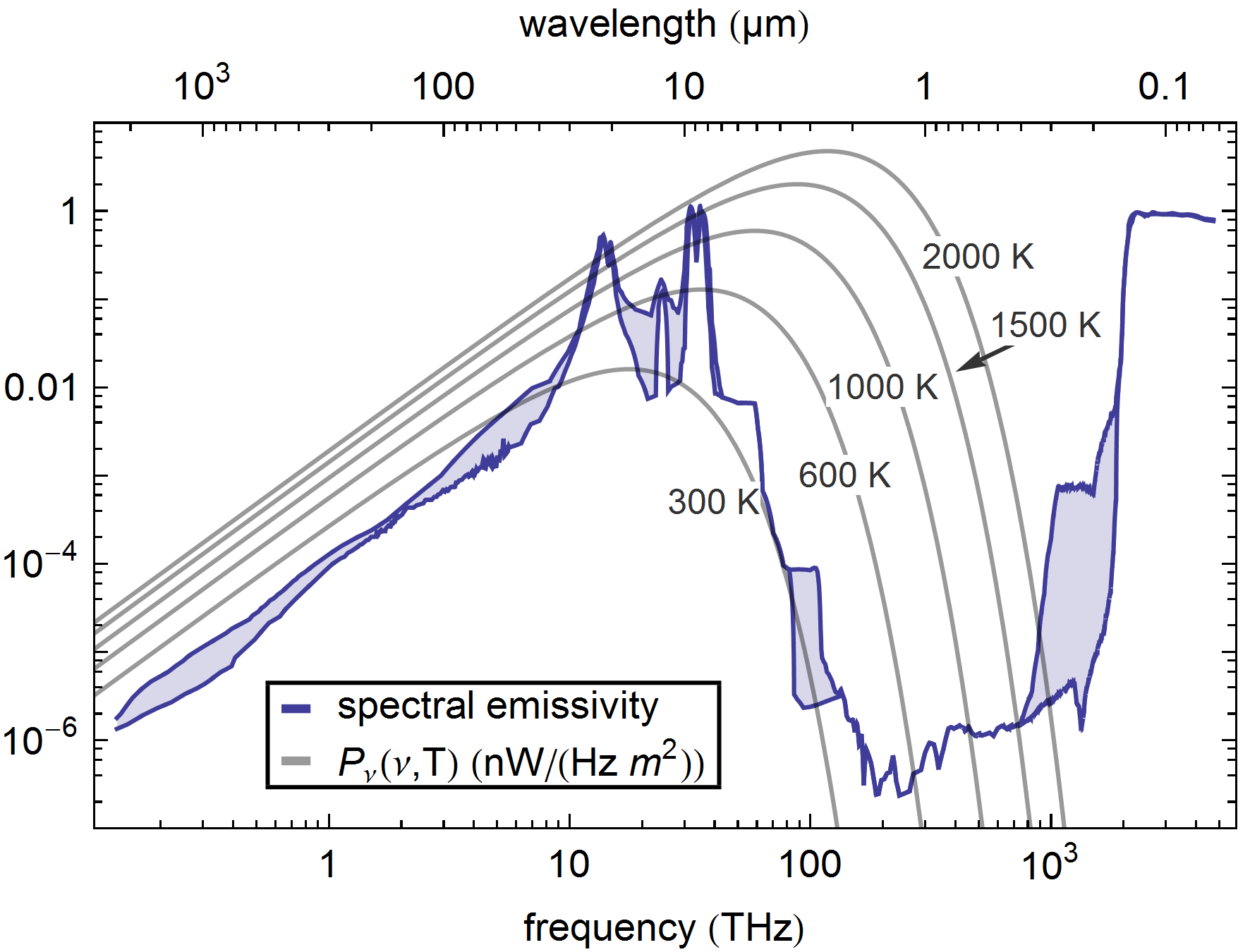}\vspace{2mm}
	\includegraphics[width=8.4cm]{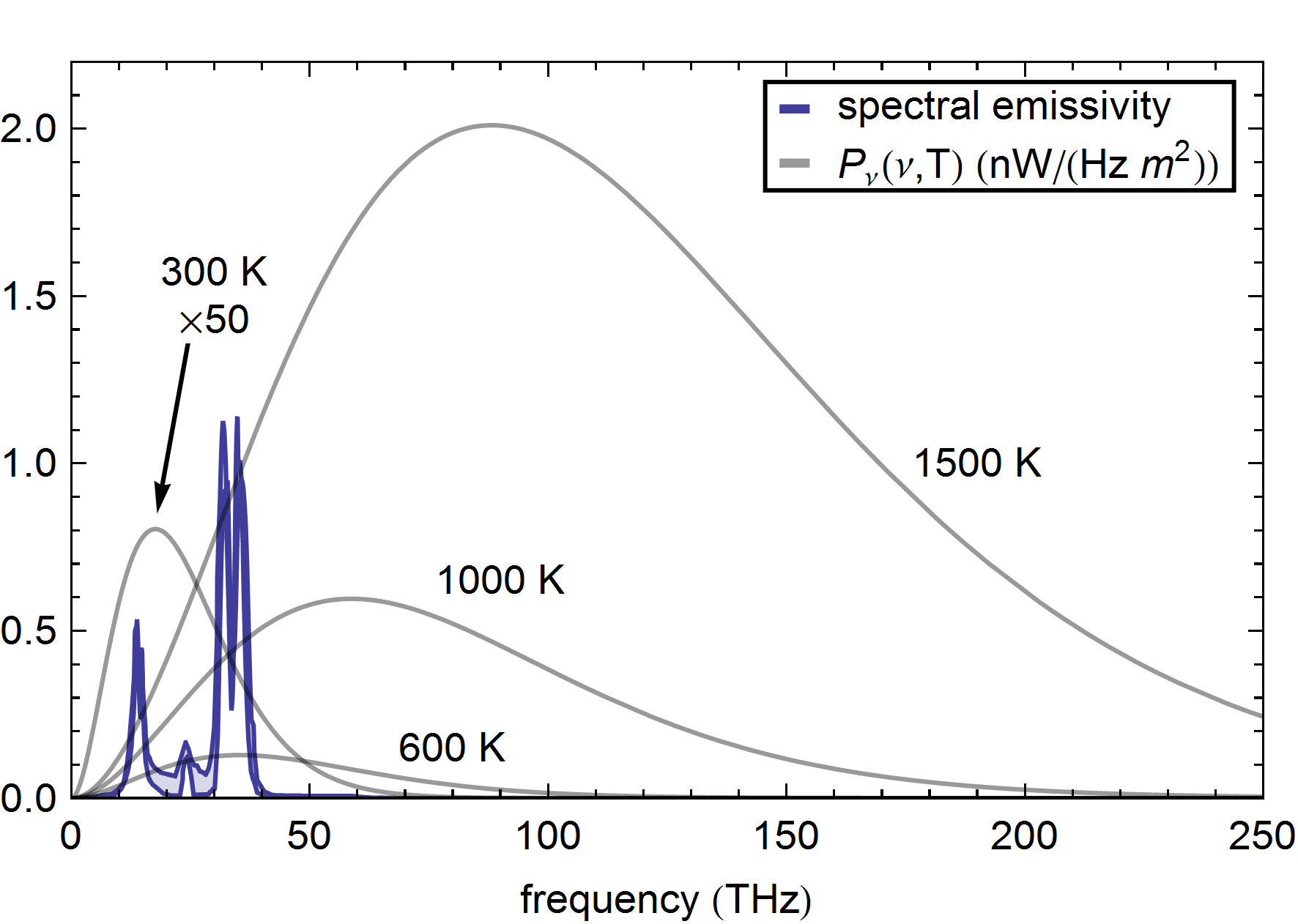}
	\caption{Spectral emissivity of the a nanofiber and black-body spectral emissive power, $P_\nu(\nu,T)$, as function of the optical frequency plotted double-logarithmically and linearly respectively. The lines delimiting the area are computed from the two curves delimiting the uncertainty region of the dielectric function, see Fig.~\ref{fig:nhat:silica}. The gray lines show the  spectral emissive power per unit surface area for the temperatures denoted within the plot. \label{fig:emissivity:fiber}}
\end{figure}

The spectral emissivity of a fused silica cylinder of 500~nm diameter is shown in Fig.~\ref{fig:emissivity:fiber}.
This graph immediately shows the cause for the different scaling of the radiated power of a nanofiber compared to a black body: With increasing temperature, a black body would emit an increasing fraction of the radiated power in the visible wavelength range. However, in this spectral range, the nanofiber is transparent and its emissivity is low. As a consequence, for the temperature range considered here, the increase of the radiated power with temperature is much lower for a nanofiber than for a black body.

\subsection*{Emissivity of a silica--vacuum interface}
\begin{figure}[h]
	\centering
	\includegraphics[width=8.4cm]{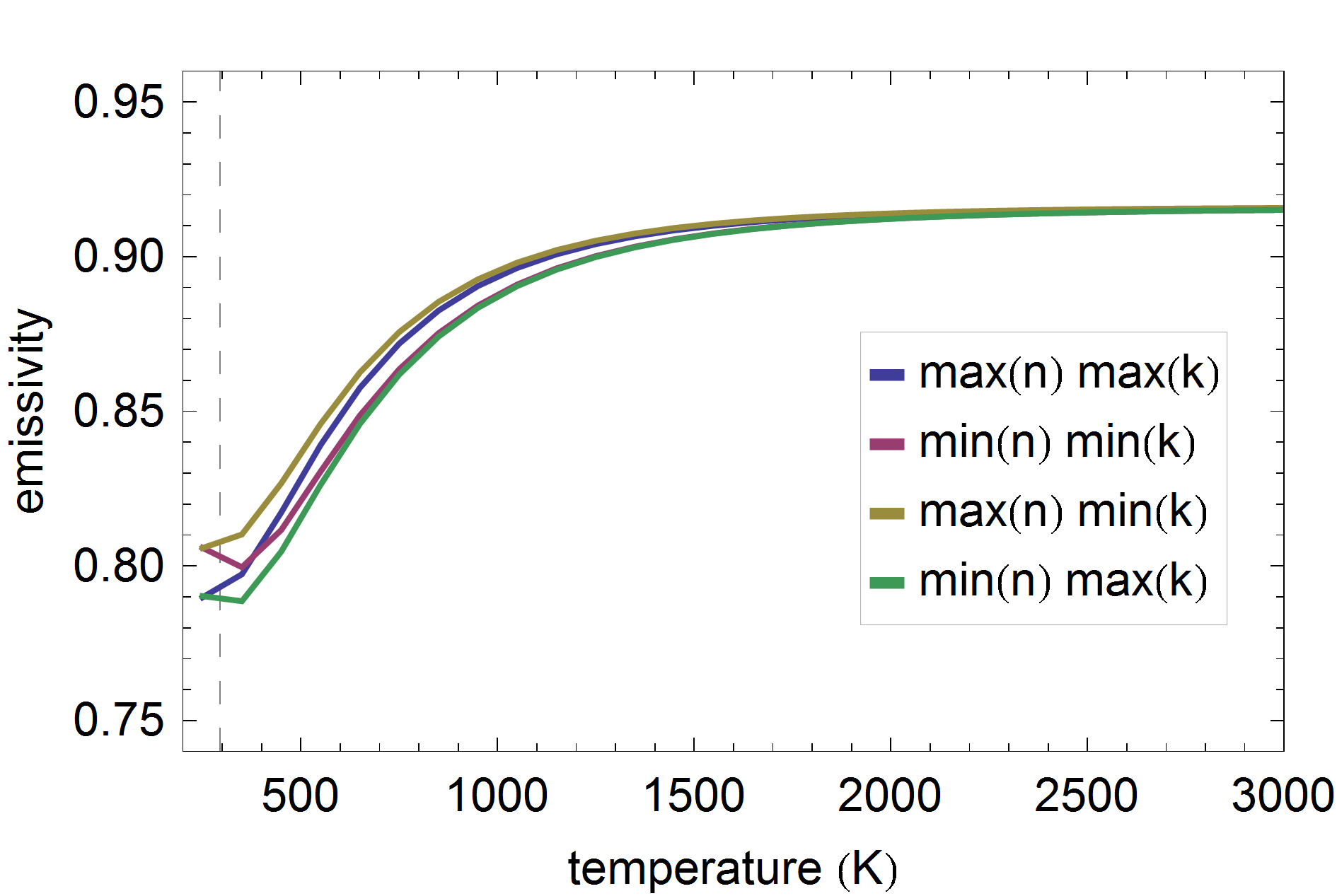}
	\caption{Emissivity of a polished silica–vacuum interface as function of temperature.}\label{fig8}
\end{figure}

We now sketch the steps that are necessary for computing the emissivity of a polished silica–vacuum interface. A comprehensive treatment of thermal radiation of surfaces can be found in the literature \citep{Siegel1972,Lienhard2011}. The total thermally radiated power of an object can be calculated by using the Stefan-Boltzmann equation $H(T)=\epsilon(T)\,\sigma_\mathrm{B}\,T^4\,A$, where $A$ is the surface area of the object, $\sigma_\mathrm{B}$ is the Stefan-Boltzmann constant, and $\epsilon (T)$ is the total hemispherical emissivity. The latter can be derived using Kirchhoff’s law, which states that the emissivity of an interface is equal to its absorptivity. For a semi-infinite silica–-vacuum interface, all light that enters the silica will be absorbed. For this reason, the directional spectral emissivity is given by $\epsilon(\nu,\Omega)=1-R(\nu,\Omega)$, where $\nu$ is the optical frequency, $\Omega=\{\theta, \phi\}$ stands for the polar and the azimuthal angle, and $R(\nu,\Omega)$ is the directional spectral reflectivity. We note that this approach neglects the transmission through the body and is thus in principle only valid for structures which are much larger than all absorption lengths in the thermal radiation spectrum. This condition is, however, not fulfilled in the case of silica at high temperatures for which the thermal spectrum significantly extends into the visible range where fused silica is highly transparent. However, in order to be able to compare to compare the FED prediction to Planck's law including a color correction that does not make any assumptions regarding the size of the object, we still pursue this approach. By using Planck’s law for the spectral radiance, see Eq.~\ref{eq:Plancks:law}, the hemispherical emissivity can then be computed:
\begin{equation}
\epsilon(T) = \frac{\int\dd \Omega\int\dd\nu\,W(\nu,T)\,\epsilon(\nu,\Omega)\,\cos(\theta)}{\sigma_\mathrm{B}T^4}~.
\end{equation}
The directional spectral reflectivity can be derived from Frensel’s formula for absorbing media by averaging over both polarizations \citep{Jackson1999}:
\begin{equation}
\begin{split}
	R(\nu,\Omega) = \frac{1}{2}\left( \left|\frac{\hat{n}_1(\nu)\cos(\theta)-\sqrt{\hat{n}_2^2(\nu)-\hat{n}_1^2(\nu)\sin^2(\theta)}}{\hat{n}_1(\nu)\cos(\theta)+\sqrt{\hat{n}_2^2(\nu)-\hat{n}_1^2(\nu)\sin^2(\theta)}}\right|^2\right. \\
	\left. +\left|\frac{\hat{n}^2_2(\nu)\cos(\theta)-\hat{n}_1\sqrt{\hat{n}_2^2(\nu)-\hat{n}_1^2(\nu)	\sin^2(\theta)}}{\hat{n}_2^2(\nu)\cos(\theta)+\hat{n}_1\sqrt{\hat{n}_2^2(\nu)-\hat{n}_1^2(\nu)\sin^2(\theta)}}\right|^2 \right )~,
\end{split}
\end{equation}
where $\hat{n}_1=1$ and $\theta$ are the complex refractive index and the angle of incidence in vacuum, respectively, and $\hat{n}_2$ is the complex refractive index of silica. 
The total hemispherical emissivity is plotted in Fig.~\ref{fig8} as a function of temperature for all combinations of the maximum and minimum values of the real and imaginary part of the refractive index.
 As explained in the main text of the manuscript, only the minimum and maximum emissivity traces (green and yellow lines, respectively) are used for further calculations.

\subsection*{Modification of the emissivity by pollutants}
The fact that the absorption of the heating laser is dominated by pollutants implies, according to Kirchhoff's law, that heat radiation in the spectral region of the heating laser wavelength is primarily emitted by the pollutants. Here we show that the overall emittance of the fiber is, however, only weakly modified by the presence of the pollutants and is well-described assuming a pure silica fiber. We first give a brief summary of the arguments which are then presented in detail.

In essence, the absorption of the pollutant in the near infrared (NIR: $\sim(0.7$--$2)~\upmu$m) spectral range only dominates because of the extreme transparency of the silica. However, it is still extremely small and, in particular, negligible compared to that of silica in the long-wavelength infrared (LWIR: $\sim(6$--$30)~\upmu$m) range where the latter is opaque, see Fig~\ref{fig:nhat:silica}. As a result, the emittance of the total structure (silica nanofiber including the pollutants) is dominated by the properties of silica in the relevant spectral regions, i.e., where the vast majority of the heat radiation is emitted.

More in detail, we determined the TOF transmission loss caused by the pollutant from a measurement of the intracavity loss at the probe wavelength and found $A\sub{pol}(\lambda\sub{probe})< 2\cdot 10^{-2}$ \cite{Wuttke2012}. Only a fraction of this transmission loss is actually caused by absorption as it also includes scattering losses due to non-adiabatic mode transformation in the taper sections and scattering losses caused by the pollutants. As a conservative estimation, we still assume that the entire transmission loss through the TOF is due to absorption caused by the pollutants. From  $A\sub{pol}(\lambda\sub{probe})$ we determine an effective extinction coefficient at the probe laser wavelength
\begin{equation}
	k\sub{eff}(\lambda\sub{probe})=\frac{A\sub{pol}(\lambda\sub{probe})\,\lambda\sub{probe}}{4\pi\,L\sub{waist} }< 3\cdot 10^{-7}~.
\end{equation}
This is seven orders of magnitude smaller than the peak extinction coefficient of fused silica in the LWIR range $k\sub{silica}\sim 1$, see Fig.~\ref{fig:nhat:silica}. 
As a consequence, even for temperatures as high as 2000~K, the spectral radiated power of the pollutants in the NIR range is much smaller than the spectral radiated power of silica in the LWIR range because the corresponding ratio of spectral radiances given by Planck's formula does by far not overcompensate this factor.

Moreover, from $A\sub{pol}(\lambda\sub{probe})$, we can infer an estimate of the effective layer thickness of the pollutants by comparing this value to what is observed in surface absorption spectroscopy of nanofiber-adsorbed dye molecules \citep{Warken2007}: If the pollutants were perfectly black molecules with an extinction coefficient $k\sub{pol}$ on the order of 1, the absorption of $A\sub{pol}(\lambda\sub{probe})=2~\%$ in the NIR spectral range would correspond to less than 1~\permil~of a closed monolayer of these molecules on the nanofiber surface or, equivalently, an effective layer thickness of less than $1$~pm. Even if $k\sub{pol}(\lambda\sub{probe})$ was two orders of magnitude lower (i.e., slightly more transparent than indium tin oxide) the corresponding effective layer thickness would still be below 0.1~nm. This means that the pollutant would contribute a volume fraction below $f\sub{V}=5\cdot 10^{-4}$ to the nanofiber which we take as an upper limit in the following. 

Using $f\sub{V}$, we can now estimate the maximum value that the effective extinction coefficient can possibly reach in any part of the spectrum. It is given by the maximum possible value of the pollutants extinction coefficient (which is on the order of $k\sub{pol}\sim 1$) multiplied by the volume fraction, yielding $k\sub{eff}(\lambda\sub{max})\sim 5\cdot 10^{-4}$. This is more than four orders of magnitude smaller than the peak value for fused silica in the LWIR range. 

\begin{figure}
	\centering
	\includegraphics[width=8.4cm]{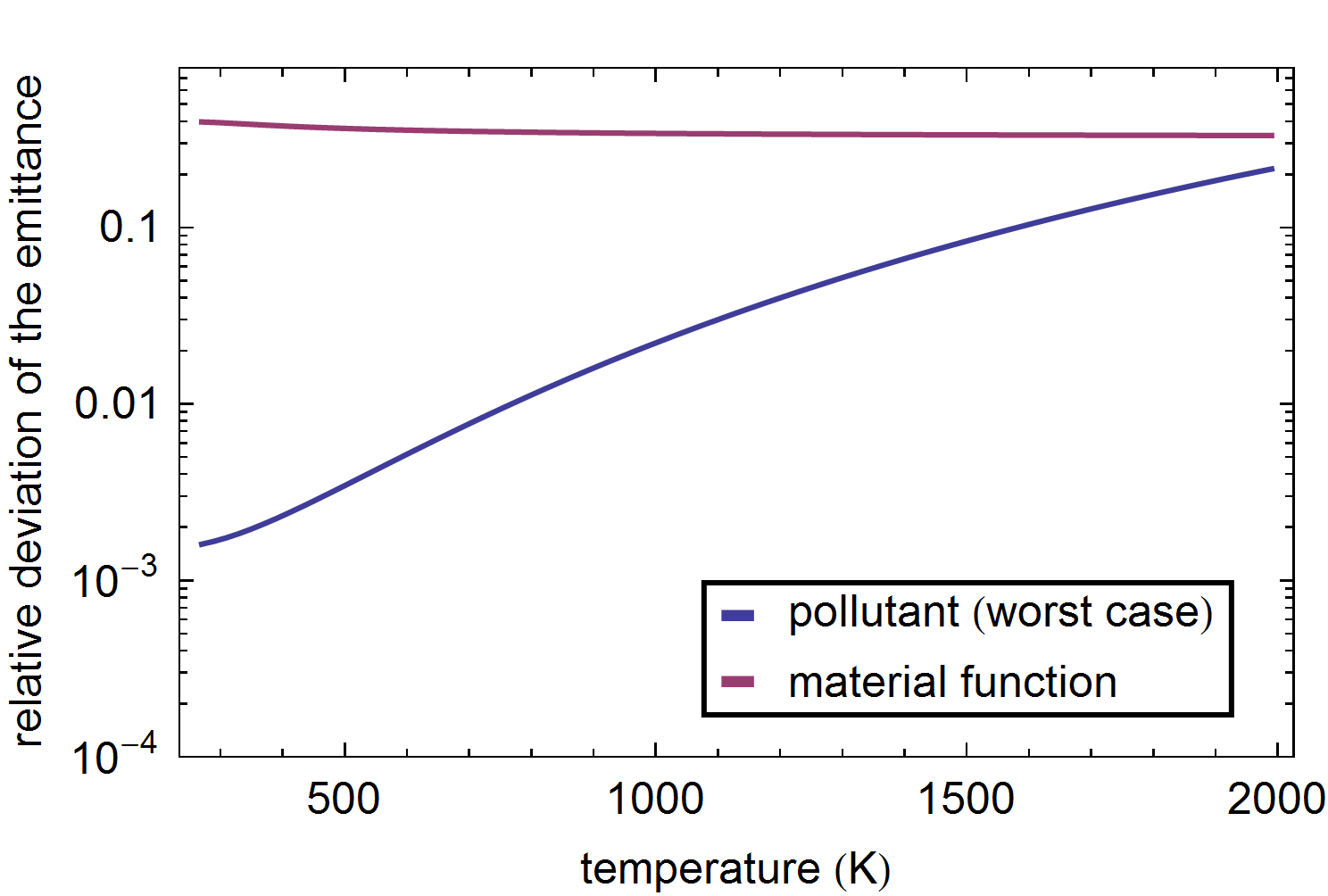}
	\caption{Calculated relative deviation of the radiated power with and without pollutant. The upper estimate of the absolute radiated power with pollutants that enters into this ratio has been obtained by adding $k\sub{eff}=5\cdot 10^{-4}$ to the spectral extinction coefficient of fused silica and computing the far-field radiation of a 500~nm cylinder composed a material with the resulting dielectric function. We also show the relative uncertainty of the emittance due to the uncertainty of the dielectric function of silica (see above). We find that the worst case estimate of the deviation caused by the pollutants is smaller than the uncertainty caused by the dieletric function. \label{fig:deviation}}
\end{figure}

Assuming this worst-case value for the entire spectrum, we find a maximum relative deviation of the radiated power with and without pollutant of 20~\% at 2000~K, see Fig.~\ref{fig:deviation}. 
We note that several worst-case estimations entered this value: We overestimated the pollutant's absorption by assuming it solely responsible for the entire TOF transmission losses. We assumed that the pollutants have a low extinction coefficient at the probe laser wavelength yielding a large layer thickness of pollutants and thus a large volume fraction. And finally, we assumed that pollutants are totally opaque ($k\sub{pol}~\sim 1$) at all wavelengths except for the probe laser wavelength.

\subsection*{Pre-heating procedure}
Prior to each cycle of heating and successive cooling, we establish a reproducible thermal situation by an initializing pulse sequence of the heating laser: First the heating laser power is set to a high value (80$-$160~mW) and is sent through the TOF for 30~s. This heating time is significantly longer than the thermalization time constant of the unprocessed fiber parts of $\sim$8~s, determined in an independent measurement by tracking the central wavelength shift of the FBG stop-band. Then, the heating laser is switched off for 2~s. This cooling time is ten times longer than the thermalization time constant of the nanofiber, see main text. This ensures that the nanofiber is thermalized at ambient temperature while the unprocessed fiber exhibits a constant initial condition.

\subsection*{Optical path length change}
The optical path length change has its origin in three basic effects: The thermo-optic and the strain-optic effect of silica, i.e., the temperature- and strain-dependence of its refractive index, $n$, as well as the radial thermal expansion of the fiber. All effects lead to a change of the so-called V-parameter of the vacuum-clad fiber waist \citep{YarivOE}, $V=2\pi a \sqrt{n^2-1}/\lambda_0$, either via a change of $n$ or via a change of the fiber radius $a$. This results in a change of the effective refractive index of the fundamental HE$_{11}$ fiber mode in which the light is guided through the TOF, $n_\mathrm{eff}=\beta(V)/k_0$, where $\beta(V)$ is the $V$-dependent propagation constant of the HE$_{11}$ mode \citep{YarivOE} and $k_0=2\pi/\lambda_0$. 

The tensile stress of the fiber varies with temperature because of the axial thermal expansion of the fiber: Following the heat-and-pull process, the TOF is pre-strained before fixing it to its mechanical mount, leading to an elongation of the nanofiber waist by $\Delta L_\mathrm{strain}/L_0=5\cdot 10^{-2}$ of its original physical length $L_0$. This elongation is much larger than the axial thermal expansion of the nanofiber which, assuming an unstrained nanofiber and a thermal expansion coefficient of silica of $\alpha_\mathrm{te}=(\Delta L/L)/\Delta T=(4-7)\cdot 10^{-7}\,/$K \citep{NISTSRM739}, would be of the order of $\Delta L_\mathrm{therm}/L_0\approx 10^{-3}$ for the maximum measured temperature of 2000~K. As a consequence, the axial thermal expansion of the nanofiber does not modify the length of the pre-strained nanofiber but rather lowers its tensile stress. This leads to a change of $n$ via the strain-optic effect and to an increase of $a$ due to the reduced transversal contraction of the nanofiber.

The thermally-induced change of the effective refractive index for a given fiber radius is thus given by: 
\begin{equation}
\begin{split}
\frac{\dd n\sub{eff}(a,T)}{\dd T} = \frac{\partial n\sub{eff}(a)}{\partial n}\left(\frac{\partial n (T)}{\partial T}-n\,\alpha\sub{te}(T)\,\alpha \sub{so}\right)\\
+\frac{\partial n\sub{eff}(a)}{\partial a} (1+\mu\sub{P})\alpha\sub{te}(T)\,a,\label{eq:dndT}
\end{split}
\end{equation}
where $\partial n(T)/\partial T$ is the thermo-optic coefficient of silica, $\alpha\sub{so}=(\Delta n /n )/(\Delta L\sub{strain}/L_0)=-0.206$ is the strain-optic coefficient of silica \citep{Borrelli1968}, and $\mu\sub{P} = -0.168$ is Poisson's ratio for silica \citep{Borrelli1968}.	

In the literature, the thermo-optic coefficient of silica is well characterized for room temperature. However, few studies have been carried out for temperatures exceeding $\sim$800 K. To our knowledge, the widest temperature range (300~K $-$ 1570~K) has been characterized using the thermally induced wavelength shift of a FBG-mirror stop-band $\Delta\lambda_\mathrm{B}$ \citep{Liao2009}. From the latter, the thermo-optic coefficient has been extracted for a wavelength of $\sim$1570~nm by using
\begin{equation}
\frac{1}{\lambda\sub{B}}\,\frac{\dd \lambda\sub{B}}{\dd T} = \left[(1-\alpha\sub{so})\alpha\sub{te}(T)+\frac{1}{n} \frac{\partial n(T)}{\partial T} \right]
\end{equation}
where $\lambda\sub{B}$ is the central wavelength of the FBG stop-band \citep{Pal2004}. Based on this data, we applied an established model that predicts the dispersion of the thermo-optic coefficient \citep{Palik1998} to compute its value at the probe wavelength of $\lambda_0=852$~nm used in our experiment. By fitting the temperature-dependent refractive index change with a 2nd-order polynomial function (coefficient of determination: $R^2=1-6\cdot 10^{-6}$) and by differentiating the latter with respect to the temperature, we obtain the temperature dependent thermo-optic coefficient:
\begin{equation}
	\frac{n(T)}{\partial T} = 9.627(9)\cdot 10^{-6}+7.74(1)\cdot 10^{-9}\,(T-299~\mathrm{K})\label{eq:therm:thermooptic:coefficient}
\end{equation} 
 We compared the prediction of our fit with values extracted from several literature sources that use various methods \citep{Wray1969,Palik1998,Tan2000,Pal2004} and find a relative error of the predicted refractive change index change with respect to the literature values of $\Delta(\partial n/\partial T)/(\partial n/\partial T)<10~\%$. We note that the data from all cited works also qualitatively agrees with the temperature dependence predicted by equation~\ref{eq:therm:thermooptic:coefficient}. We also note that, to our knowledge, the thermal expansion coefficient is known up to 1000~K \citep{NISTSRM739}. For higher temperatures we therefore employ a linear extrapolation of the two highest temperature values measured in \citep{NISTSRM739}.

\begin{figure}[h]
	\centering
	\includegraphics[width=8.4cm]{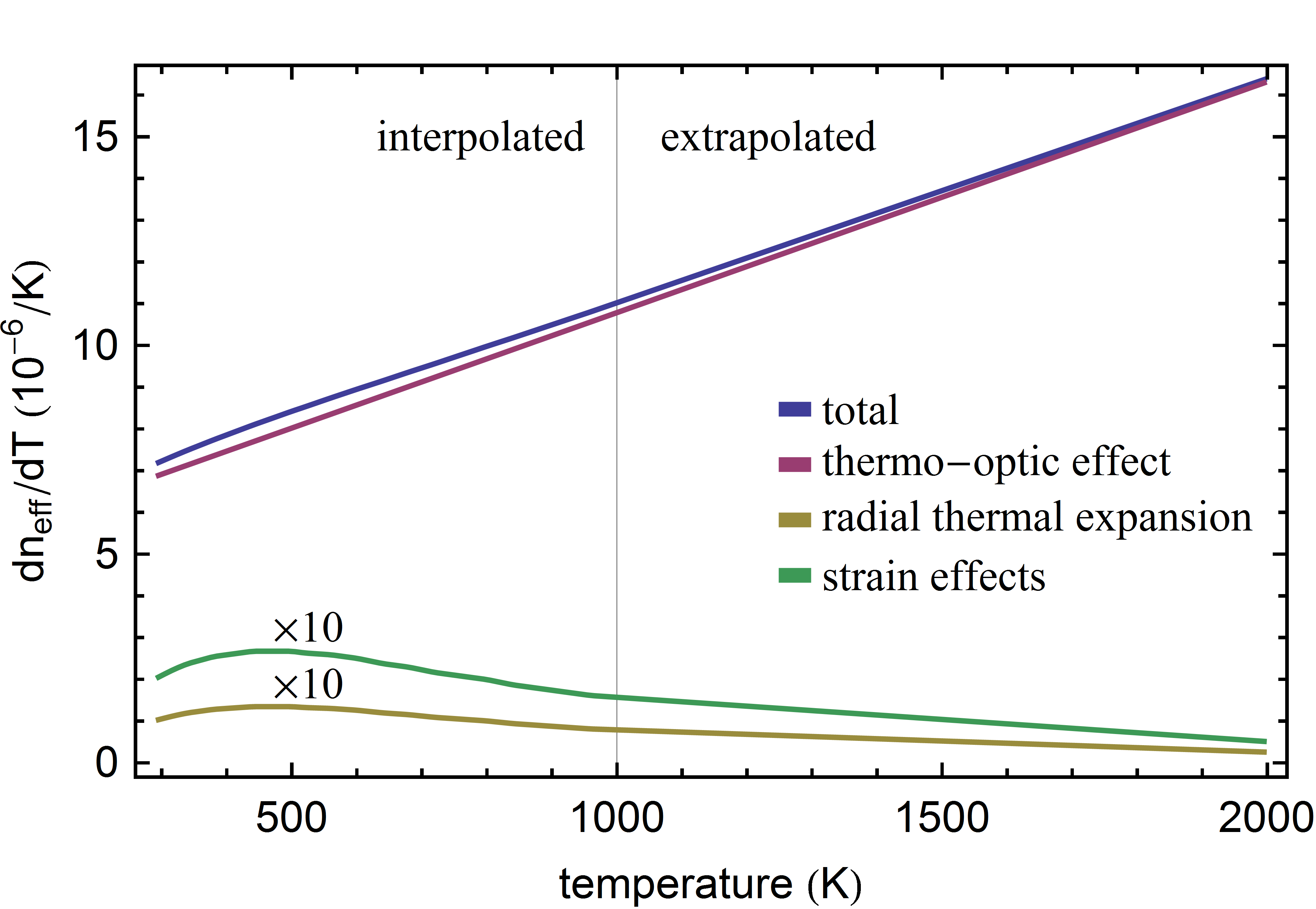}
	\caption{Temperature dependence of the effective thermo-optic coefficient and its contributions for a fiber diameter of 500~nm.}\label{fig4}
\end{figure}

The resulting thermal change in the effective refractive index is plotted as function of temperature in Fig.~\ref{fig4}. 
In all the above considerations, the effect of the fiber core has been neglected. This is justified because the core constitutes a negligible fraction ($<0.5~\%$) of the fiber material which contributes to the guiding of the light in the nanofiber section and, in addition, has similar optical properties as the cladding material.

\subsection*{The TOF radius profile }
\begin{figure}[htb]
	\centering
	\includegraphics[width=8.4cm]{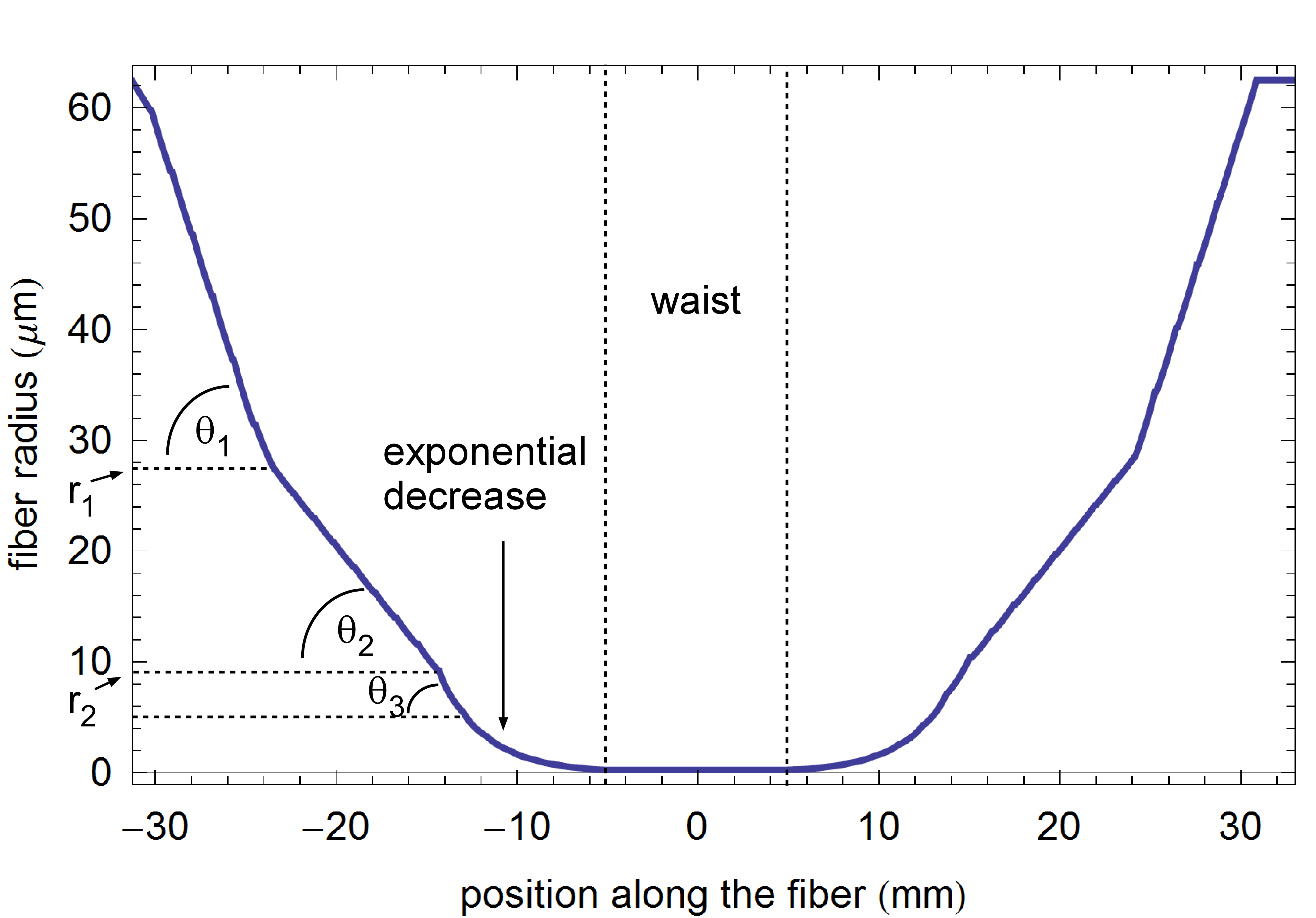}
	\caption{Simulated radius profile of the tapered section of TOF-resonator \#1.}\label{fig5}
\end{figure}

The tapered optical fibers used in this experiment are produced from standard single-mode optical fibers with a cladding diameter of 125~$\mu$m. For this purpose, we employ a custom-built computer-controlled fiber pulling rig which allows us to fabricate TOFs with a predetermined radius profile. We chose a TOF radius profile which includes a nanofiber waist with a constant subwavelength diameter that is enclosed symmetrically by two fiber tapers as shown in Fig.~\ref{fig5}.
In a standard optical fiber, the light is guided by the core–cladding interface whereas in the nanofiber section the core has a negligible effect and the light is guided the cladding–vacuum interface. In these subwavelength optical nanofibers, a significant fraction of the optical power propagates outside of the fiber in the form of an evanescent wave that extends several hundred nanometers into the vacuum. A standard fiber exhibits a small refractive index difference between the core and the cladding and the fundamental fiber mode can be approximated by the weakly guided LP$_{01}$-mode. To reach a high transmission through the TOF, this mode has to be adiabatically transformed by the taper transition to match the fundamental HE$_{11}$-mode of the strongly guiding nanofiber \citep{YarivOE}. This is achieved by a taper profile that is composed of three linear sections followed by an exponential decrease adjacent to the waist (see Fig.~\ref{fig5} for TOF-resonator \#1). In the first and third linear section of the taper, the fiber diameter is reduced with comparably large angles, $\Theta_1$ and $\Theta_3$, whereas in the second linear section, the transition from a core- to cladding-guided mode takes place and the angle $\Theta_2$ has to be chosen small enough to ensure an adiabatic transformation. 

The TOFs used in the experiment are produced from Fibercore SM800 single mode optical fibers, using the parameters given in Tab.~\ref{tab1}, where $r_1$ and $r_2$ are the cladding radii at the transitions between the linear taper sections. 
\begin{table}
	\begin{tabular}{|c|c|c|c|c|}
		\hline
		sample & $\Theta_1$ & $\Theta_1$ & $\Theta_1$ & $L_\mathrm{waist}$\\\hline
		TOF-resonator \#1 & 5 mrad & 2 mrad & 4 mrad & 10 mm\\\hline
		TOF-resonator \#2 & 5 mrad	& 2 mrad & 4 mrad & 5 mm\\\hline
	\end{tabular}
	\caption{Parameters of the two TOF radius profiles of the two samples used in this experiment.}\label{tab1}
\end{table}
Previous measurements have shown that the actual TOF radius profile agrees with the simulated radius profile within the measurement accuracy of $\approx 10$~\% \citep{Wiedemann2010,Stiebeiner2010}.

\subsection*{Mechanical stability at high temperatures}
While the upper temperature limit exceeds the glass transition temperature of fused silica of about 1450~K \citep{Doremus2002} no breaking due to melting has been observed. To investigate this further we heated TOF-resonator~\#2 to its breaking point by increasing the heating power step-wise. Here, the pre-heating procedure has been omitted to prevent it from damaging the fiber and the heating time was extended to $1.25$~s. We found a maximum heating power at which the fiber maintained its mechanical stability as high as $192(10)$~mW with $\Delta L_\mathrm{opt}^\mathrm{max}=235(5)~\upmu$m, the latter corresponding to a predicted equilibrium waist temperature of $T\sub{waist} = (2515\pm 255)$~K. We note, that the uncertainty only includes that of the heat exchange rates and is actually higher due to the extrapolation used for the thermo-optic coefficient. 

\begin{figure}[htb]
	\centering
	\includegraphics[width=8.4cm]{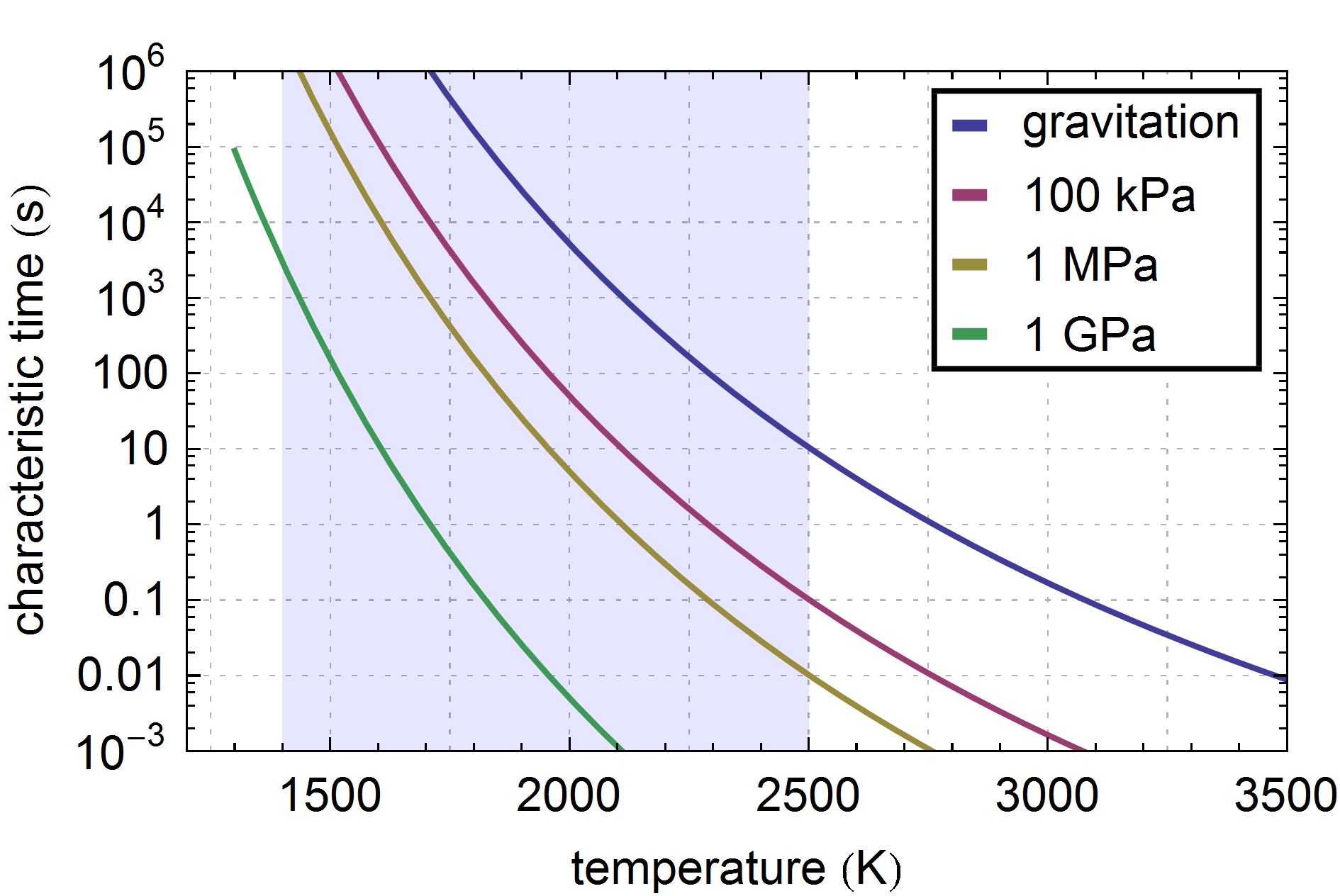}
	\caption{Characteristic time scales of the viscous deformation dynamics caused by gravitation and different axial strains. The interpolated range is marked blue. \label{fig:chartimes}}
\end{figure}

We now theoretically examine the mechanical stability of the fiber by treating it as a thin viscous filament with a temperature dependent viscosity $\eta(T)$. Therefore, we compute the characteristic time of the viscous dynamic in dependence of the temperature and compare it to the measurement cycles. As the fiber is pre-strained we have to consider two different characteristic time constants: The relaxation from the initial strain in the fiber $\sigma_i$ by elongation which happens on the timescale of 
\begin{equation}
	\tau_s=\frac{3\,\eta}{\sigma_i}
\end{equation} 
and the fusing at which the initial strain is already released and gravity causes the deformation 
\begin{equation}
	\tau_v = \frac{\eta}{3\,\rho\,g\,L_0}~,
\end{equation}
where $g$ is the gravitational acceleration \citep{Koulakis2008}. We use an Arrhenius-model fit to the viscosity of silica from literature which was obtained from measurement values in the temperature range $(1400-2500)$~K: 
\begin{equation}
	\eta(T) = 5.8\cdot 10^{-8}\,\exp\left(E\sub{a}/(RT)\,\right)\mathrm{~Pa\,s}~,
\end{equation}
 where $R$ is the molar gas constant and $E_a=515.4~\mathrm{kJ/mol}$ is the activation energy \citep{Doremus2002}. The results for the characteristic times are shown for the cases in which the fiber only experiences gravity and that of axial strain in Fig.~\ref{fig:chartimes}. 
We find a rapid decrease of the characteristic times with increasing temperature that covers many orders of magnitude. When the fiber is heated, the strain is released due to viscous elongation of the fiber up to the point at which the strain is too small for a significant change on the time scale of one measurement cycle $\tau_s > 60$~s. With that we estimate the remaining stress after heating to temperature of $T\approx 1800$~K to be $\sigma> 100$~kPa. As this is the stress at the maximum temperature and, thereby, maximum thermal elongation, it is ensured that the fiber stays stressed during the measurements.

The temperature at which the fiber fuses should occur when the characteristic time $\tau_v\approx(0.5-5)$~s becomes comparable to the measurement cycle which occurs at a temperature of $T=(2710\pm140)$~K. This is in excellent agreement with the value obtained from the measurement

%

\end{document}